\documentstyle[12pt]{article}
\newtheorem{Def}{Definition}
\newtheorem{The}[Def]{Theorem}
\newtheorem{Pro}[Def]{Proposition}
\newtheorem{Lem}[Def]{Lemma}

\newtheorem{Rem}{Remark}
\newtheorem{Exa}{Example}
\newtheorem{Alg}{Algorithm}

\newcommand{\N}{{\rm I\!N}}
\newcommand{\Z}{{\rm Z\!\!Z}}
\def\r#1{{\rm r}_{#1}}
\def\val#1{{\rm val}_{#1}}

\begin{document}
\title{Numeration systems on a regular language}
\date{March 3, 1999}
\author{Pierre B. A. Lecomte and Michel Rigo\\
Institut de Math\'{e}matiques, Universit\'{e} de Li\`{e}ge, \\
Grande Traverse 12 (B 37), B-4000 Li\`{e}ge, Belgium.\\
{\tt plecomte@ulg.ac.be}, {\tt M.Rigo@ulg.ac.be}}

\maketitle
\begin{abstract}
Generalizations of linear numeration systems in which $\N$ is recognizable by finite automata are obtained
by describing an arbitrary infinite regular language following the lexicographic ordering. For these
systems of numeration, we show that ultimately  periodic sets are recognizable. We also study the
translation and the multiplication by constants as well as the  order-dependence of the recognizability.
\end{abstract}
\section{Introduction}
A series of recent papers are devoted to numeration systems \cite{Ber,BH,Dur1,Dur2,Fr,Ho,Sha} 
and are mainly concerned with the study of the so-called recognizable sets of integers. Roughly speaking,
a set of integers is recognizable if their representations have a very simple syntax,
i.e. if they form a regular language.

An usual way of representing integers, leading to the so-called linear representation systems,  is to
consider a strictly increasing sequence $(U_n)_{n\in \N}$  of integers and to use some algorithm (such as
the greedy algorithm) to represent each natural number  $x$  by a word $c_0\ldots c_n$ such that $c_0 U_n +
\cdots + c_n U_0 = x$ \cite{Fra}. For example, with $U_n = p^n$ and the greedy algorithm, one gets the
standard numeration system with basis
$p$. 

Among the sets of integers possibly recognizable, $\N$ is of special interest. For instance,  if it is
recognizable, then one  can easily check whether a word over the alphabet of the digits represents an
integer or not. Under quite general assumptions, it is shown in \cite{Sha} that for $\N$ to be recognizable, it
is necessary that $U_n$ satisfies a linear recurrence relation.  The sufficient condition given in \cite{Ho}
is that $U_n$ satisfies an extended beta polynomial  for the dominant root
$\beta>1$ of the recurrence . Examples of such systems are the numeration systems defined by a recurrence relation whose
characteristic polynomial  is the minimum polynomial of a Pisot number (like the standard numeration systems or the Fibonacci
system \cite{BH}).

A nice description of the recognizable sets has been obtained for the latter \cite{BH,BHMV}. They are the sets of integers
that can be defined in the Presburger arithmetic extended by some predicate related to the considered Pisot number. In
particular, various operations do preserve the recognizability, such as addition, translation and multiplication by a
constant.

In \cite{Co}, Cobham shows that the only sets that are simultaneously recognizable with respect to two standard numeration
systems having multiplicatively independent integer basis are precisely the finite unions of arithmetic progressions. This
remarkable result has been extended to non standard linear systems \cite{BHMV,Dur1,H,PB}, the more general version being
obtained quite recently in \cite{Dur2}.

In the above mentionned results, a property of the considered systems seems to play a crucial role: 
the representation $x\in\N\mapsto r(x)\in\{digits\}^*$ is increasing with respect to the lexicographic ordering (this is an
assumption in \cite{Sha}, it is a consequence of the greedy algorithm). Observe that a numeration system having this property is
completely determined by the language $r(\N)$ and the ordering of the digits, the sequence $U_n$ and the algorithm defining the
system being just extra data devised to compute the function $r$ in some ``practical" fashion. 

Taking this into account, we thus define an {\it (abstract)  numeration  system} as being a triple $S=(L,\Sigma,<)$ where $L$
is an infinite language over the totally ordered alphabet $(\Sigma, <)$. Enumerating the elements of $L$ lexicographically with
respect to $<$ leads to a one-to-one map $\r{S}$ from $\N$ onto $L$. To any natural number $n$, it  assigns the $(n+1)^{th}$ word
 of $L$, its {\it $S$-representation}, while the reciprocal map $\val{S}$ sends any word belonging to $L$ onto its {\it numerical
value}. A subset $X\subset\N$ is said to be {\it $S$-recognizable} if $\r{S}(X)$ is a regular 
subset of $L$.

Having in mind a possible generalization of the Cobham's theorem, it is natural to check whether the ultimately periodic subsets of
$\N$ are $S$-recognizable . Of course, if they are, then $L=\r{S}(\N)$ is regular. It is a quite remarkable fact that conversely,
if $L$ is regular, then every arithmetic progression is indeed $S$-recognizable (a special case of this result has been obtained
separately in \cite{M}). 

As recalled above, the recognizability of $\N$ is an important property that is often required. Unless
otherwise stated, we  assume in the sequel that $L$ is a regular language. Under this assumption, we obtain algorithms to
compute $\r{S}$ and $\val{S}$. The first is a generalization of the greedy algorithm involving the complexity functions of
the states $w^{-1}. L$, $w\in\Sigma^*$, of the minimal automaton of $L$ in place of the sequence $U_n$  (for more about minimal
automaton, see for instance \cite{E}) . Both proved to be quite usefull in many concrete experiments.

In a positional numeration system, each digit has its own weight so that the question of changing the order of the digits is
somewhat irrelevant in this case. In an abstract numeration system, the letters has no {\it a priori} individual role and, as we
show with the help of the language ${\{a,b\}^*\setminus a^*b^*}$, the family of recognizable sets depends on the ordering of the
alphabet. However, we exhibit two classes  of regular languages for which the recognizability of a set of integers is independent
of the order  on the alphabet. One of these classes is the set of the slender languages \cite{ADPS}.  The other is the set
of the languages $L\subset
\Sigma^*$ for which the complexity functions of the associated languages $w^{-1}. L$ differ only at finitely many places.

As for the stability of the recognizability under natural arithmetic operations, we show that for each $t$, a subset $X$ of
$\N$ is $S$-recognizable if and only if $X+t$ is $S$-recognizable. On the other hand, multiplication by a constant generally does
not preserve recognizability so that addition is not a regular map as well. For example, in the numeration system $S$ based on the
language $a^* b^*$, the set of $t\in\N$ for which $tX$ is $S$-recognizable if $X$ is $S$-recognizable consists of the perfect squares.
Note that in this case, the function $\val{S}$ is nothing else but the well known Peano's function \cite{Rus}; surprisingly, the proof
of the result is difficult and it relies partly on the properties of the Pell's equation \cite{Ari1,Ari2}.
\section{Basic definitions and notations}
In this paper, if $\Sigma$ is a finite alphabet then $\Sigma^*$ is the free monoid (with identity $\varepsilon$) 
generated by $\Sigma$. For a set $S$, $\# S$ denotes the cardinality of $S$ and for a string 
$w\in \Sigma^*$, $|w|$ denotes the length of $w$.

Let $L\subset \Sigma^*$ be a regular language. We denote $M_L=(K,s,F,\delta,\Sigma)$ the minimal automaton of $L$ where $K$ is
the set of states, $s$ is the initial state, $F$ is the set  of final states and $\delta:K\times
\Sigma \to K$ is the transition function.  We often write $k.\sigma$ instead of $\delta(k,\sigma)$.  

 Recall that the elements of $K$ are the sets ${w^{-1}.L=\{ v \in \Sigma^* \, : \, wv \in L \}}$, $w\in\Sigma^*$. The state $k$
is of the form $w^{-1}.L$ if and only if $k=s.w$, $w^{-1}.L$ being then the set $L_k$ of words accepted by $M_L$ from $k$. In
particular, $L=L_s$.

We denote $u_l(k)$ the number $ \# (L_k \cap \Sigma^l)$ of words of length $l$ belonging to $L_k$
and $v_l(k)$ the number of words of length at most $l$ belonging to $L_k$, 
$$v_l(k) = \sum_{i=0}^l u_i(k).$$
If we are only interested in the number of words belonging to $L$, then we simply note $u_l$ and $v_l$ instead 
of $u_l(s)$ and $v_l(s)$ provided that it does not lead to any confusion.

\begin{Def}{\rm
A {\it numeration system} $S$ is a triple $(L,\Sigma,<)$ where $L$ is an infinite regular language over the 
totally ordered alphabet $(\Sigma,<)$.

For each $n\in\N$, $\r{S}(n)$ denotes the $(n+1)^{th}$  word of $L$  with respect to the
lexicographic ordering and is called the {\it $S$-representation} of $n$.

Remark that the map $\r{S}: \N \to L$ is an increasing
bijection. For $w\in L$, we set $\val{S} (w)=\r{S}^{-1} (w) $. We call $\val{S}(w)$ the {\it numerical value} of $w$.
}\end{Def}

\begin{Def}{\rm 
Let $S$ be a numeration system. A subset $X$ of $\N$ is {\it $S$-recognizable} if $\r{S} (X)$ 
is recognizable by finite automata.
}\end{Def}

Let $S=(L,\Sigma,<)$ be a numeration system. Each $k\in K$ for which $L_k$ is infinite leads to the 
numeration system $S_k=(L_k,\Sigma,<)$. The applications $\r{S_k}$  and $\val{S_k}$ are simply denoted $\r{k}$ and $\val{k}$ if the
context is clear. If $L_k$ is  finite, the applications $\r{k}$ and $\val{k}$ are defined as in the infinite case but the domain of the
former restricts to ${\{0,\ldots ,\# L_k -1\}}$.
\section{Computation of $\val{S}$ and recognizability of ultimately periodic sets}
In this section, given any numeration system
$S=(L,\Sigma,<)$, we indicate how to compute the function $\val{S}$ and show that the arithmetic progressions  $p+\N\,
q$  are $S$-recognizable. 

\newpage
We first need a lemma.

\begin{Lem} \label{lemme}
Let $S=(L,\Sigma,<)$ be a numeration system. If $\alpha \beta$ belongs to $L_k$, $\alpha,\beta \in \Sigma^+$, 
then
$$\val{k}(\alpha \beta)=\val{k.\alpha} (\beta)+v_{|\alpha \beta|-1} (k)-v_{|\beta|-1} (k.\alpha)
+\sum_{{\alpha'<\alpha}\atop{|\alpha'|=|\alpha|}} u_{|\beta|} (k.\alpha').$$
\end{Lem}
{\it Proof.} We have to compute the number of words belonging to $L_k$ and lexicographically strictly lesser 
than $\alpha\beta$. There are three kinds of such words. The first consists of words of length strictly lesser 
than $\alpha\beta$ and counts $v_{|\alpha\beta|-1} (k)$ elements. The next one consists of words of length $|\alpha\beta|$ admitting
the prefix $\alpha$. Since a word $\alpha' \beta'$ belongs to $L_k$ if and only if $\beta'$ belongs to $L_{k.\alpha'}$, we see that
there is $\val{k.\alpha} (\beta) - v_{|\beta|-1} (k.\alpha)$ such words. It is clear that there is 
$$\#\{ w \in L_k \, : \, w=\alpha' \beta', |\alpha'|=|\alpha|, |\beta'|=|\beta|\ {\rm and}\ 
\alpha'<\alpha \} = \sum_{{\alpha'<\alpha}\atop{|\alpha'|=|\alpha|}} u_{|\beta|} (k.\alpha')$$
words of the last kind. $\Box$

\begin{Rem}
{\rm Taking for $\alpha$ a letter in lemma \ref{lemme} one would deduce easily an effective algorithm to compute $\val{S}$.} 
\end{Rem}

\begin{Rem}{\rm It follows also from lemma \ref{lemme} that for each word $w$, 
$$\val{S}(w) = \sum_{{k\in K} \atop{0 \le l < |w|}}c_{k, l} \, u_l(k) $$ 
where the ``digits" $c_{k, l}$ are less or equal to $\# \Sigma$.
}\end{Rem}

\begin{The} \label{pa}
Let $S=(L,\Sigma,<)$ be a numeration system and $p$, $q$ two non negative integers. The arithmetic progression 
$p+ \N \, q$ is $S$-recognizable.
\end{The}
{\it Proof.} We can assume that $p<q$. We show that the  the minimal automaton of ${\cal A}=\r{S} (p+ \N \,
q)$  is finite. Its  states are the sets
$$w^{-1}.{\cal A} = \{ x \in \Sigma^* \, :\, \val{S}(wx) \equiv p\ {\rm mod}\ q \}, w\in \Sigma^* .$$ 
Observe first that the sequence $v_n(s)$ being a solution of a linear recurrence equation, is ultimately periodic in $\Z_q$, say of
period $t$. By lemma
\ref{lemme}, for $|w|$ large enough, $w^{-1}.{\cal A}$ is thus of the form
$$\{ x\, :\, \val{k}(x)  + v_{|x|+i}(s) - v_{|x|-1}(k) + \sum_{k'\in K} j_{k'}\, u_{|x|}(k') \equiv p\ {\rm mod}\ q \}$$ 
for some $k \in K$, $j_{k'} \in \{ 0, \ldots , q-1 \}$ and $i\in \{ 0, \ldots , t-1 \}$. $\Box$
\section{Computation of $\r{S}$ and reordering of the alphabet}
We now explain how to compute effectively $\r{S}$ and discuss to what extend the $S$-recognizable subsets of $\N$ depend on the
ordering of the alphabet.

Let $S=(L,\Sigma,<)$ be a numeration system, where ${\Sigma = \{\sigma_1 < \cdots < \sigma_p\}}$.

It is clear that  
$$|\r{S} (n)| = \inf_m \{ m \, |\, n<v_m \}.$$ 
Set $|\r{S} (n)|=l$ then $n-v_{l-1}$ is the number of words of length $l$ belonging to $L$ 
and strictly lesser than $\r{S} (n)$. 
 
To determine the first letter of the representation, we have to compute the number $N_t^l$ of words of length $l$ 
belonging to $L$ and begining with $\sigma_1$ or $\ldots$ or $\sigma_t$ ($t\le p$)
$$N_t^l=\sum_{i=1}^t u_{l-1} (\sigma_i^{-1} .L).$$
If $N_{t-1}^l \le n-v_{l-1}<N_t^l$ then the first letter is $\sigma_t$. We proceed in the same way to 
find out the other letters of the representation. Recall that if $k$ is a state of $M_L$ 
then $\delta(k,\sigma_j)=\sigma_j^{-1}.k$. Hence, the following algorithm that computes the $S$-representation $w$ of a given integer $n$. 

\begin{Alg}\label{algorithme} 
Let $l$ such that $v_{l-1} \le n < v_l$,\\
$k \leftarrow s$\\
$m \leftarrow n-v_{l-1}$\\
$w \leftarrow \varepsilon$\\
for $i$ ranging from $1$ to $l$ do\\
\indent $j \leftarrow 1$\\
\indent while $m\ge u_{l-i} [\delta(k,\sigma_j)]$ do\\
\indent\indent $m \leftarrow m-u_{l-i} [\delta(k,\sigma_j)]$\\
\indent\indent $j \leftarrow j+1$\\
\indent $k \leftarrow \delta(k,\sigma_j)$\\
\indent $w \leftarrow w\sigma_j$.
\end{Alg}

\begin{Rem} {\rm If $\lim\limits_{n\to +\infty} \frac{v_{n+1}}{v_n}=\theta < \infty$ then the 
temporal complexity of the algorithm is ${\cal O}((\#\Sigma) \log_\theta n)$.} \end{Rem}

As an easy application of algorithm \ref{algorithme}, we obtain a first class of numeration systems in which the recognizable sets
are independent of the order of the alphabet.  

It is convenient to introduce notations for the change of numeration
systems. Given systems $S=(L,\Sigma,<)$ and $T=(L',\Sigma',\prec)$, we set 
$$\Theta_{S,T} = \r{T} \circ \val{S} : L\to L' \ {\rm and}\ \Theta_{S,T}' = \val{T} \circ \r{S} : \N \to \N.$$
If the underlying $S$ and $T$ are known from the context, we simply write $\Theta$ and $\Theta'$.

\begin{Pro}\label{classe} 
Let $S=(L,\Sigma,<)$ and $T=(L,\Sigma,\prec)$ be two numeration systems. 
Let $n_0$ be a non negative integer. If for all states $k$ and $k'$ of $M_L=(K,s,F,\delta,\Sigma)$, 
$$u_n (k)= u_n (k'),\ \forall n\ge n_0,$$
then $X\subset \N$ is $S$-recognizable if and only if $X$ is $T$-recognizable.
\end{Pro}
{\it Proof.} Assume that $\Sigma=\{\sigma_1< \cdots <\sigma_p\}=\{\sigma_{\nu_1}\prec\cdots\prec\sigma_{\nu_p}\}$ where $\nu$ is a
permutation of $\{1, \ldots ,p \}$.  We prove that the graph $\hat{\Theta}=\{ (x,y) \in L \times L : \val{S} (x) = \val{T} (y) \}$ of $\Theta$
is regular over the alphabet
$\Sigma \times \Sigma$, showing that $\r{T}(X) = p_2(\hat{\Theta} \cap p_1^{-1}(\r{S}(X))$ is regular if and only if $\r{S}(X)$ is
regular, where $p_1, p_2 :(\Sigma \times \Sigma)^* \to \Sigma^*$ are the canonical homomorphisms of projection. 

Let $(x,y)$ belonging to $\hat{\Theta}$. The two systems $S$ and $T$ have the same sequence $(v_n)_{n\in \N}$, 
thus ${|x|=|y|}$. 

By algorithm \ref{algorithme}, if $|x|\ge n_0$ then 
$$x =\underbrace{ \sigma_{i_1} \ldots \sigma_{i_l}}_{\alpha} \beta \ {\rm and} \  y = \underbrace{\sigma_{i_{\nu_1}} \ldots
\sigma_{i_{\nu_l}}}_{\alpha'}\beta'$$ 
where $|\beta|= |\beta'|= n_0$, $\beta \in L_{s.\alpha}$, $\beta' \in L_{s.\alpha'}$ and
$$\val{S_{s.\alpha}}(\beta) = \val{T_{s.\alpha'}}(\beta').$$
To conclude, it is then sufficient to observe that the words of
$\hat{\Theta}$ of length at least $n_0$ are exactly the words accepted by the following nondeterministic finite automaton.
The set of states is $(K \times K) \cup \{ f \}$. The initial state is $(s,s)$. The new symbol $f$ denotes the unique final state.
According to what precedes, there are two kinds of transitions. First those of label $(\sigma_i, \sigma_{\nu_i})$ mapping the state $(k,k')$
onto $(k.\sigma_i,k'.\sigma_{\nu_i})$. 
Second those of label $(\beta,\beta')$ mapping $(k,k')$ onto $f$, provided that $|\beta|= |\beta'|= n_0$, $\beta \in L_{k}$, $\beta'
\in L_{k'}$ and $\val{S_{k}}(\beta) = \val{T_{k'}}(\beta')$. $\Box$ 

\begin{Exa} The language over the alphabet $\{ a,b \}$ consisting of the 
words containing an even number of $a$ satisfies the hypothesis of proposition \ref{classe}.
\end{Exa}

In the next proposition, we give equivalent formulations of the assumption of proposition \ref{classe}. They are expressed in terms of the
incidence matrix $A_L$ of the minimal automaton $M_L$ of $L$. Recall that it is the matrix defined by
$$(A_L)_{i,j}= \sum_{t=1}^p \delta_{k_i.\sigma_t , k_j},\ 1\le i,j \le \kappa,$$ 
where the $\sigma_t$'s and the $k_i$'s denote the $p$ letters and the $\kappa$ states of $M_L$ respectively.

We denote $f_L$ the characteristic vector of the set of final states:
$$(f_L)_i=\left\{\begin{array}{ll}1 & {\rm if}\ k_i \in F\cr 0 & {\rm otherwise}.\cr \end{array}\right.$$ 
Observe that
\begin{equation} \label{A-f}
(A_L^m f)_i =\sum_{j=1}^{\kappa} (A_L^m)_{i,j} f_j = u_m (i).
\end{equation}

\newpage
\begin{Pro} Let $L$ be a regular language over an alphabet $\Sigma$ and 
$M_L=(K,s,F,\delta , \Sigma)$ be its minimal automaton. Let $m$ be the multiplicity of $0$ as 
root of the minimum polynomial of $A_L$.  Let $r>m$. The next assertions are equivalent
\begin{enumerate}
\item $\forall n\ge r$, $\forall k,k' \in K$, $u_n (k)= u_n (k')$,
\item $\forall n\ge m$, $\forall k,k' \in K$, $u_n (k)= u_n (k')$,
\item $\exists \lambda \in \N_0$ : $A^m_L f = \lambda\, v$, with $v=(1,\ldots ,1){\mbox{\~{ }}}$.
\end{enumerate}
In particular, $\forall k \in K$, $\forall i\ge 0$, $u_{m+i} (k) = (\#\Sigma)^i u_m (k)$.
\end{Pro}
{\it Proof.} This follows immediately from (\ref{A-f}) and the well known fact that any polynomial that is cancelled out by $A_L$ is the
characteristic polynomial of a linear recurrence equation satisfied by each of the sequences $u_n(k)$. $\Box$
\medskip

Here is another easy characterization of the languages for which the assumption of proposition \ref{classe} holds true.

\begin{Pro} Let $L$ be a regular language over an alphabet $\Sigma$. It satisfies the hypothesis of proposition 
\ref{classe} if and only if there exist $n_0,u_0\in \N$ such that for all $w \in \Sigma^*$, $\# ((w^{-1}.L) \cap \Sigma^{n_0}) = u_0$. $\Box$
\end{Pro}

The set of slender languages is the second class of languages for which the recognizable sets of integers are 
independent of the ordering of the alphabet.

\begin{Def}{\rm \cite{ADPS} 
Let $d$ be a positive integer. The language $L$ is said to be {\it $d$-slender} if
$$\forall n\ge 0,\ u_n (s)\le d,$$
$L$ is said to be {\it slender} if there exists $d$ such that $L$ is $d$-slender.
}\end{Def}

\begin{Lem}\label{ptlemme} 
{\rm \cite{Sha}} Let $L$ be a regular language over the totally ordered alphabet $(\Sigma,<)$. 
The set ${\cal I}(L,<)$ (resp. ${\cal G}(L,<)$) obtained by taking from all the words of $L$ of the same length only the first (resp. last) one in the 
lexicographic order is regular. $\Box$
\end{Lem}

\begin{Pro} 
Let $d$ be a positive integer. Let $L$ be a regular $d$-slender language. 
Let ${S=(L,\Sigma,<)}$ and ${T=(L,\Sigma,\prec)}$ be two numeration systems. 
If $X\subset \N$ is $S$-recognizable then $X$ is $T$-recognizable.
\end{Pro}
{\it Proof.} Like in the proof of Proposition \ref{classe}, we show that the graph $\hat{\Theta}$ of the change of systems is regular. Using
lemma \ref{ptlemme}, we define iteratively the regular languages $I_{i,<}$ and $I_{i,\prec}$ by
$$\left\{ \begin{array}{lcl}
I_{1,<} & = & {\cal I}(L,<) \cr
I_{1,\prec} & = & {\cal I}(L,\prec),
\end{array} \right.$$ 
and, for $i=2,\ldots ,d$, 
$$\left\{ \begin{array}{lcl}
I_{i,<} & = & {\cal I}[L\setminus (\bigcup\limits_{j=1}^{i-1} I_{j,<}),<] \cr
I_{i,\prec} & = & {\cal I}[L\setminus (\bigcup\limits_{j=1}^{i-1} I_{j,\prec}),\prec ].
\end{array} \right.$$
Since for all $x\in L$, $|x|=|\Theta(x)|$, the graph of $\Theta$ is thus given by
$$\hat{\Theta} = \bigcup_{j=1}^d \left[ (I_{j,<} \times I_{j,\prec}) \cap (\Sigma \times \Sigma)^* \right].\ \Box $$ 

In spite of the two previous propositions,  the change of ordering of the alphabet generally does not preserve the recognizability as  we 
shall see about $\Sigma=\{a,b\}$ and $L=\Sigma^*\setminus a^*b^*$.

\begin{Lem}\label{chgt}  
Let $n \in \N$. For $U=(\Sigma^*,\Sigma,a < b)$ and
$V=(\Sigma^*,\Sigma,b \prec a)$ one has
$$\Theta_{U,V}' (n) = 3.2^{l}-n-3,$$ 
where $l=|\r{U} (n)|.$
\end{Lem} 
{\it Proof.} Observe that since $\# \Sigma^l = 2^l$,  if $w_1 < \cdots < w_{2^l}$ then $w_{2^l} \prec \cdots \prec w_1$.  Moreover  
$2^l-1\le n \le 2^{l+1}-2$. Thus 
$$\Theta' (n) = 2^{l+1}-2-[n-(2^l-1)].\ \Box$$

\medskip

\begin{Pro} Let $\Sigma=\{a,b\}$ and $L=\Sigma^*\setminus a^*b^*$.
For all $n\ge 2$, if $l=|\r{U} (n-1)|$ then
$$\Theta_{S,T} (ba\, b^n)=ab\, a^{n-l-1} \, b \, \r{U} (n-1),$$
where $S=(L,\Sigma,a<b)$, $T=(L,\Sigma,b \prec a)$ and
$U=(\Sigma^*,\Sigma,a<b)$. In particular, $\val{S}(ba\, b^2 b^*)$ is not $T$-recognizable.
\end{Pro}
{\it Proof.} The minimal automaton $M_L$ of $L$ is given by
\begin{center}
\begin{picture}(200,70)(0,0)
\put(0,40){\circle{15}}
\put(-2,38){\shortstack{s}}
\put(100,40){\circle{15}}
\put(98,37){\shortstack{t}}
\put(200,40){\circle{15}}
\put(200,40){\circle{11}}
\put(197,38){\shortstack{p}}
\put(10,40){\vector(1,0){80}}
\put(48,43){\shortstack{$b$}}
\put(110,40){\vector(1,0){80}}
\put(148,43){\shortstack{$a$}}
\put(-20,40){\vector(1,0){10}}
\put(0,20){\oval(8,24)}
\put(7,18){\shortstack{$a$}}
\put(100,20){\oval(8,24)}
\put(107,18){\shortstack{$b$}}
\put(200,20){\oval(8,24)}
\put(207,18){\shortstack{$a,b$}}
\end{picture}\\
Figure 1. The minimal automaton of $\Sigma^*\setminus a^*b^*$.
\end{center}
Therefore $L_p=\Sigma^*$, 
$$\left\{ \begin{array}{ll}
u_0 (s) = u_1 (s) = 0, & \cr
u_n (s) = 2^n-n-1, & \forall n \ge 2,\cr
\end{array}
\right.$$
while $u_n (t) = 2^n-1$ for all $n \in \N$.

In $L$, there are $v_{n+1}(s)$ words of length at most $n+1$, 
$u_{n+1}(s)$ words of length $n+2$ begining with $a$ and 
$u_n (p)-1$ words of length $n+2$ begining with $ba$. 
Hence, the number of words belonging to $L$ and lexicographically lesser than $ba\, b^n$ is
$$\val{S} (ba\, b^n)=\sum_{i=2}^{n+1} (2^i-i-1) + 2^{n+1} + 2^n - n -3.$$

Using lemma \ref{lemme}, 
we sketch the computation of $\val{T} [ab\, a^{n-l-1} \, b \, \r{U} (n-1)]$
$$\begin{array}{cl}
= & \val{t} [a^{n-l-1} \, b \, \r{U} (n-1)] + \sum\limits_{i=2}^{n+1} (2^i-i-1)  + 2^n + n \\
= & \val{p} [a^{n-l-2} \, b \, \r{U} (n-1)] + \sum\limits_{i=2}^{n+1} (2^i-i-1)  + 2^{n+1} -1 \\
\vdots & \\
= & \val{p} [b \, \r{U} (n-1)] + \sum\limits_{i=2}^{n+1} (2^i-i-1)  + 2^{n+1} -1 + \sum\limits_{i=l+2}^{n-1} 2^i \\
= & \val{p} [\r{U} (n-1)] + \sum\limits_{i=2}^{n+1} (2^i-i-1)  + 2^{n+1} -1 + \sum\limits_{i=l+2}^{n-1} 2^i + 2^l \\
= & \Theta' (n-1) + \sum\limits_{i=2}^{n+1} (2^i-i-1)  + 2^{n+1} -1 + 2^n - 3.2^l. 
\end{array}$$
Hence the value of $\Theta_{S,T} (ba\, b^n)$, in view of lemma \ref{chgt}. Applying the pumping lemma, it is now straightforward to check that $\val{S}(ba\, b^2 b^*)$ is
not $T$-recognizable. $\Box$

\section{Translation by a constant}
Here we show that the $S$-recognizability of a set is conserved under the translation by a constant. 
First  we recall some classical results about numeration systems.

\begin{Lem} \label{Norm}
{\rm \cite{Fr}} Let $p \in \N \setminus \{ 0,1\}$. The normalization function 
$$\nu : \{ 1,\ldots ,p \}^* \to \{ 0,\ldots ,p-1 \}^*$$
which gives the normalized representation in base $p$ of an integer 
(the representation obtained by the greedy algorithm) is a rational function, its graph $\hat{\nu}$ 
is recognizable by a finite letter-to-letter automaton. $\Box$
\end{Lem}

\begin{Lem} \label{Defin}
{\rm \cite{BHMV}} A subset of $\N$ is recognizable in base $p\ge 2$ if and only if it is definable in 
the structure $\langle \N , + , V_p \rangle$, where for $x\neq 0$, $V_p (x)$ is the greatest power of $p$ dividing $x$ while 
$V_p (0)=1$. $\Box$
\end{Lem}

\begin{Pro} \label{Trans}
Let $S=(L,\Sigma,<)$ be a numeration system. For each natural number $t$, 
 $X+t$  is $S$-recognizable if $X \subset \N$ is $S$-recognizable.
\end{Pro}
{\it Proof.} Let $\Sigma =\{ \sigma_1 < \cdots < \sigma_p \}$ and let the homomorphism $h:\Sigma^* \to \{1,\ldots ,p\}^*$ be defined
by $h : \sigma_i \mapsto i$. For $x \in \N$, the word ${h(\r{S}(x))=x_0 \ldots x_l \in \{1,\ldots ,p\}^*}$ 
is a representation in base $p$ of the integer ${\pi_p (h(\r{S}(x))) = x_0 \, p^l+ \cdots + x_l \, p^0}$. 

Since $L$ is regular over $\Sigma$, by lemma \ref{Norm}, $\nu (h(L))$
is regular over ${\{0,\ldots,p-1\}}$ and by lemma \ref{Defin}, the set
$${\cal N} = \pi_p [\nu (h (L))] $$
is definable in $\langle \N , + , V_p \rangle$. 

The successor function $S_L:L \to L$ (with respect to the lexicographic order) is then
regular.  Indeed, ${\cal S} =  \pi_p \circ \nu \circ h \circ S_L\circ(\pi_p \circ \nu \circ h)^{-1}$ is the restriction to $\cal N$ 
of the fucntion $x\mapsto y$ defined in $\langle \N , + , V_p \rangle$ by the formula 
$$(y \in {\cal N}) \wedge  (x<y) \wedge (\forall z) (z \in {\cal N} \wedge x<z)\to (y \le z).$$ 

Assume now that $X$ is $S$-recognizable, i.e. that $\r{S}(X)$ is a regular set. Then  ${\r{S}(X+t) = S_L^t(\r{S}(X))}$ is regular. $\Box$
\section{Multiplication by a constant}

In this section, we show that, in general, the multiplication by a constant does not preserve the recognizability. To that end, we use the
system $S = (a^*b^*,\{a,b\} ,a<b)$, for which it is easy to see that
$$\val{S} (a^p b^q)=\frac{1}{2} (p+q)(p+q+1)+q.$$ 

\begin{Rem} {\rm Observe that the r.h.s.
is nothing else but the well-known Peano's function \cite{Rus}.}
\end{Rem}

It would suffice to show that, say, the multiplication by two does not preserve recognizability but here we are lucky enough to get
more.

\begin{The} \label{Mult}
Let $S$ be the numeration system $(a^*b^*,\{a,b\} ,a<b)$ and let $\alpha \in \N$. The multiplication by $\alpha$ transforms
the $S$-recognizable sets into $S$-recognizable sets if and only if $\alpha$ is a perfect square.
\end{The}
{\it Proof.} (i){\it Sketch.} If $\alpha$ is not a perfect square, we show that for a suitably choosen $r$,
$${\cal L}_\alpha^r = a^r b^* \cap \r{S} (\alpha \val{S}(a^*))$$
is infinite while the set of lengths $|{\cal L}_\alpha^r|$ only contains finite arithmetic progressions so that 
$\r{S} (\alpha \val{S}(a^*))$  is not even context free, thanks to Parikh's theorem \cite{Par}. 

If $\alpha=\beta^2$, $\N^2$ is divided into $\beta+1$ regions $R_i$ in each of which 
an explicit formula for the function $M:(p,q)\mapsto (r,s)$ such that ${\alpha\, 
\val{S}(a^p b^q)=\val{S}(a^r b^s)}$ can be supplied. These regions come from length considerations: given a word of length $l$ and of
numerical value
$x$,  there is $\beta+1$ possible lengths for the word of value $\alpha x$. The fact that the multiplication by
$\alpha$ preserves the regularity of the subsets of $a^*b^*$ follows then from an easy lemma.

\medskip
\noindent
(ii) {\it Case of a non perfect square.} Let $\alpha$ be a non perfect square integer. We have
$$l\in |{\cal L}_\alpha^r| \Leftrightarrow \exists p :\, \val{S} (a^r b^{l-r}) = \alpha\, \val{S} (a^p).$$
In other words, $l\in |{\cal L}_\alpha^r|$ if and only if 
\begin{equation} \label{eqdepart}
 [2(r+s)+3]^2-\alpha (2p+1)^2 = 8 r+9-\alpha
\end{equation} for some $p$, where $s=l-r$. 

To guarantee that $|{\cal L}_\alpha^r|$ be infinite, we choose $r$ in such a way that

\begin{equation} \label{eqpell}
X^2-\alpha Y^2=8 r+9-\alpha
\end{equation} 
has infinitely many solutions with odd components.
To that purpose, it suffices to choose $r$ such that $8 r+9-\alpha >0$ and that the equation (\ref{eqpell})
admits a solution $(x,1)$ with $x$ odd (cf. Appendix). This can be achieved with $r$ of the form $z^2$. Indeed, the equation $x^2 - 8 z^2 =9$
has infinitely many solutions given by
$$\pmatrix{x_0\cr z_0}=\pmatrix{3\cr 0} ,\ 
\pmatrix{x_{i+1}\cr z_{i+1}} = \pmatrix{3 & 8\cr 1& 3}\pmatrix{x_i\cr z_i} ,\ \forall i\in \N.$$
The $x_i$'s are odd. We choose $i$ such that $8 z_i^2+9-\alpha >0$ and take $x=x_i$.

The set of the solutions of (\ref{eqpell}) with odd components is a finite union of sequences 
${(X_n^{(j)},Y_n^{(j)})_{n\in \N}}$, $j=1,\ldots ,m$, such that $X_n^{(j)} > C^n$ for some $C>1$ (cf. Appendix).

We are now in position to show that $|{\cal L}_\alpha^r|$ only contains finite arithmetic progressions. Suppose to the contrary that it
contains an infinite progression. Then there exist $\lambda, \mu \in \N,\mu > 0,$  and, for each $t \in \N$, indices $n_t \in \N,
j_t \in \{1, \dots , m\}$ such that 
$$\lambda +\mu t = X_{n_t}^{(j_t)} > C^{n_t}. $$
Given $t$, the sequence $n_0, \dots , n_{mt}$ contains at least $t$ distincts numbers. Therefore
$$\forall t \in \N,\  \lambda + \mu mt > C^t ,$$
a contradiction.

\noindent
(iii){\it The case of a perfect square.} Let $\alpha=\beta^2$ and $\beta$ be an odd integer. The case $\beta$ even is treated in the same
way.

We want to compute $r, s$ such that $\alpha\, \val{S}(a^pb^q) = \val{S}(a^r b^s)$, i.e.
$$[2(r+s)+3]^2-\beta^2 [2(p+q)+3]^2=8 r-8 p  \beta^2 - 9 (\beta^2-1).$$
Let $l=p+q$, $l'=r+s$. Then
$$\alpha  l(l+1)\le 2\alpha  \val{S}(a^p b^q) \le\alpha  l(l+3)  \ {\rm and}\ 
l'(l'+1)\le 2 \val{S}(a^r b^s) \le l'(l'+3).$$
Therefore, $l'(l'+1)\le\beta^2 l(l+3)$ and $\beta^2 l(l+1)\le l'(l'+3)$. 
From this, it follows easily that
$$r+s = \beta (p+q) + \left\lfloor \frac{\beta}{2} \right\rfloor + i$$
and thus
$$\left \{ \begin{array}{l}
r=r_i(p,q):=\beta (i+1)p-\beta (\beta-i-1)q+\frac18[(\beta +2i+2)^2-9] \cr
s=s_i(p,q):=-\beta i p + \beta(\beta -i)q-\frac18 [(\beta +2i)^2 -9] -1 \cr
\end{array}\right.$$
for some $i \in \{ -1,\ldots ,\beta-1 \}$. These equations together with the conditions $r, s \ge 0$ define
$\beta+1$ regions $R_i$ which divide $\N^2$.

The regular subsets of $a^*b^*$ are the finite unions of sets of the form 
$$D=\{a^{y+fz}b^{w+gx}:f,g\ge 0\},$$
$w,x,y,z \ge 0$. Substituting $y+fz$ and $w+gx$ in place of $p$ and $q$ respectively in  $r_i(p,q)$ and $s_i(p,q)$, one sees that 
$D'=\r{S} [\alpha\, \val{S}(D\cap R_i)]$ is of the form (\ref{ineg}) of lemma \ref{lemineg} below, the matrix $A$ being 
$$A=\pmatrix{z\beta (i+1) & -x\beta (\beta-i-1) \cr -z\beta i & x \beta (\beta -i)\cr}.$$
One can apply the lemma to see that $D'$ is regular except if $i=-1$ or $xz=0$. In these cases, $D'$ is easily shown to be regular by direct
inspection. $\Box$

\newpage
\begin{Lem}\label{lemineg}
Let $A$ be a non singular $p \times p$ integral matrix. For $ i = 1, \dots ,p$, set
$$h_i({\bf n}) = A_{i1}n_1 + \cdots + A_{ip}n_p -b_i,$$
where ${\bf n} = (n_1, \dots ,n_p) \in \N^p$ and $b_1,\dots,b_p \in \Z$. If the entries of ${\rm dtm} (A)A^{-1}$ are non negative,
then  the language
\begin{equation}\label{ineg}
{\cal L} = \{ a_1^{h_1} \dots a_p^{h_p}: h_1({\bf n}) \ge 0 , \dots ,h_p({\bf n}) \ge 0, {\bf n} \in \N^p \}
\end{equation}
is a regular subset of $a_1^* \dots a_p^*$.
\end{Lem}
{\it Proof.} If ${\bf n} \in \N^p$ satisfies $h_i({\bf n})\ge 0$ then $(A{\bf n})_i=b_i+ u_i$, i.e.
\begin{equation}\label{entier}
n_i = \sum_{j=1}^p (A^{-1})_{ij} (b_j + u_j),
\end{equation}  for some $u_i \in \N$.

We need to describe those ${\bf u} = (u_1,\dots,u_p)\in \N^p$ for which (\ref{entier}) defines non negative integers $n_i$.

If ${\rm dtm} (A)<0$, the entries of $A^{-1}$ are negative, there are finitely many such ${\bf u}$ and ${\cal L}$ is finite. 
If ${\rm dtm} (A)>0$, $(A^{-1})_{ij} \ge 0$, for large enough $u_j$'s, (\ref{entier}) defines thus positive numbers $n_i$ 
but it remains to ensure that they are integers. To that purpose, since $A^{-1}={\cal A}/{\rm dtm} (A)$,
where the entries of ${\cal A}$ are natural numbers, it is necessary and sufficient that the remainders 
$r_j \in \{0,\dots, {\rm dtm} (A)-1\}$ of the division of $u_j$ by ${\rm dtm} (A)$ satisfy $$\sum_{j=1}^p {\cal A}_{ij} (b_j + r_j) \equiv 0 \ {\rm \ (mod \ dtm}(A){\rm )}. $$ 
There is a finite number of such $(r_1,\dots,r_p)$ so that ${\cal L}$ is a finite union of regular languages of the form
 $$\left(a_1^{{\rm dtm} (A)}\right)^*a_1^{s_1{\rm dtm} (A)+r_1} \dots \left(a_p^{{\rm dtm} (A)}\right)^*a_p^{s_p{\rm dtm} (A)+r_p}.$$ 
(The $s_j$'s are choosen to guarantee that the $u_j$'s be large enough for the corresponding $n_i$'s to be non negative.)  $\Box$


\section{Appendix}

a) The next proposition sumarizes the well known facts that are used in he proof of theorem \ref{Mult}. The reader will find in \cite{Ari1,Ari2} the material necessary 
to achieve its proof.

\begin{Pro}Assume that $\alpha \in \N$ is not a perfect square and that $N>0$ is a natural number.

\noindent
(i) The set of solutions $(X,Y)\in \N^2$ of the equation $X^2-\alpha Y^2=N$ is the (finite) union of the sequences $(X_n,Y_n)_{n\in \N}$ defined by 
\begin{equation} \label{eqrec}
\pmatrix{X_{i+1}\cr Y_{i+1}} = \pmatrix{u & \alpha v\cr v& u}\pmatrix{X_i\cr Y_i} ,\ \forall i\in \N, \ {\rm and}\  0<X_0 \le u\sqrt{N},
\end{equation}
where $(u,v)\in \N^2$ is the minimal non trivial solution of $U^2 - \alpha V^2 = 1$, i.e. that for which $u>1$ is the smallest.

\noindent 
(ii) Each component of any solution $(X_n,Y_n)_{n\in \N}$ of (\ref{eqrec}) are solutions of $$ Z_{i+2}= 2u Z_{i+1}-Z_i,\ \forall i\in \N.$$ 
In particular, $X_{2n}, X_{2n+1},Y_{2n}$ and  $Y_{2n+1}$ are of the same parity as $X_{0}, X_{1},Y_{0}$ and  $Y_{1}$ respectively.

\noindent
(iii) For any solution $(X_n,Y_n)_{n\in \N}$ of (\ref{eqrec}), one has $X_n > u^n$. $\Box$

\end{Pro}

\noindent
b) Taking advantage of lemmas \ref{Norm} and \ref{Defin}, we give another proof of theorem \ref{pa}, based on the notion of substitution.
\begin{Lem} \label{psub} {\rm \cite{BH,BHMV,Co2}}
A subset $X$ of $\N$ is recognizable in base $p$ if and only if the characteristic sequence of $X$ is 
generated by a $p$-substitution. $\Box$
\end{Lem}

\begin{Lem} \label{trans} {\rm \cite{Co2}}
The set of the infinite words generated by $p$-substitution is closed under finite transduction. $\Box$
\end{Lem}
{\it Proof of theorem \ref{pa}.} We use the notations of proposition \ref{Trans}. The set $\nu (h(L))$ is a 
regular subset of $\{ 0 , \ldots ,|\Sigma|-1 \}^*$ and by lemma \ref{psub}, the characteristic sequence $\Psi$ of 
$\pi_{|\Sigma|} [\nu (h(L))]$ is generated by a $|\Sigma|$-substitution. To conclude, use lemma \ref{trans} and 
observe that the characteristic sequence of $\pi_{|\Sigma|} [\nu (h(p+\N\, q))]$ is the image of $\Psi$ 
under the following finite transducer (the tail has $p$ nodes and the head counts $q$ of them)
\begin{center}
\begin{picture}(235,140)(0,0)
\put(-18,70){\vector(1,0){10}}
\put(0,70){\circle{15}}
\put(9,70){\vector(1,0){23}}
\put(11,75){\shortstack{$1/0$}}
\put(40,70){\circle{15}}
\put(49,70){\vector(1,0){23}}
\put(51,75){\shortstack{$1/0$}}
\put(73,67){\shortstack{$\cdots$}}
\put(89,70){\vector(1,0){23}}
\put(91,75){\shortstack{$1/0$}}
\put(120,70){\circle{15}}
\put(127,77){\vector(1,1){20}}
\put(120,90){\shortstack{$1/1$}}
\put(153,103){\circle{15}}
\put(162,103){\vector(1,0){23}}
\put(164,108){\shortstack{$1/0$}}
\put(193,103){\circle{15}}
\put(200,96){\vector(1,-1){20}}
\put(210,90){\shortstack{$1/0$}}
\put(226,70){\circle{15}}
\put(219,63){\vector(-1,-1){20}}
\put(210,45){\shortstack{$1/0$}}
\put(193,37){\circle{15}}
\put(165,34){\shortstack{$\cdots$}}
\put(153,37){\circle{15}}
\put(146,44){\vector(-1,1){20}}
\put(120,45){\shortstack{$1/0$}}
\end{picture}\\
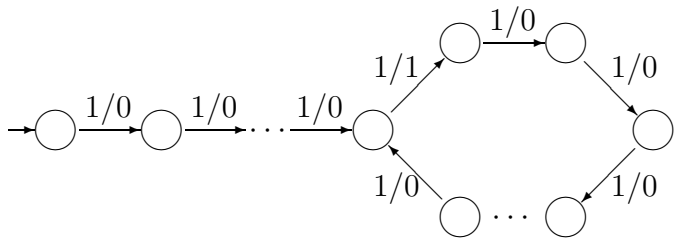
Figure 2. The finite transducer for $\pi_{|\Sigma|} [\nu (h(p+\N\, q))]$.
\end{center}
each state has a loop which corresponds to the reading and the writing of $0$. $\Box$
\medskip

\noindent
c) The nature of the $S$-recognizable sets seems to depend strongly on the system $S$.
In standard numeration systems with integer basis, the set of squares is not recognizable \cite{BHMV}
while an example of system for which it is recognizable may be found in \cite{M}, p. 141. Here is another example, based on lemma \ref{ptlemme}.

\begin{Pro}
Let $S=(a^*b^*\cup a^*c^*,\{ a,b,c\} ,a<b<c)$. The set $\{n^2 : n\in \N\}$ is $S$-recognizable. 
\end{Pro}
{\it Proof.} Indeed, since $\# ((a^*b^*\cup a^*c^*) \cap \Sigma^n) = 2n+1$, the greatest word of length $n$ in $a^*b^*\cup a^*c^*$ has numerical value $n^2$. $\Box$

\medskip
Using the same idea, one can easily produces various examples of unusual recognizable sets, such as $\{v_n: n \in \N \}$ for any regular language $L$.
\section*{Acknowledgments}
The authors would like to thank V. Bruy\`{e}re for her valuable advices and encouragements and M. De Wilde for fruitful comments.

\end{document}